\documentclass[preprintnumbers,amsmath,amssymb]{revtex4}
\usepackage{graphicx}
\usepackage{bm}
\def\be{\begin{equation}}
\def\ee{\end{equation}}
\def\bee{\begin{eqnarray}}
\def\eee{\end{eqnarray}}
\begin{document}

\author{Giuseppina Modestino}
\email{modestino@lnf.infn.it}


\affiliation{%
INFN, Laboratori Nazionali di Frascati, via Enrico Fermi 40, I-00044, Frascati (Roma) Italy\\
}%

\date{\today}
\title{Orbital velocity}

\date{\today }

\begin{abstract}

The trajectory and the orbital velocity are determined for an object moving in a gravitational system, in terms of fundamental and independent variables. In particular, considering a path on equipotential line, the elliptical orbit is naturally traced, verifying evidently the keplerian laws. The case of the planets of the solar system is presented. 
\end{abstract}

\maketitle

\section{Introduction}
\label{intro}
Although there are a myriad of treatises dealing the velocity of an object passing trough a gravitational field, we can hardly find procedures or expressions in terms of independent and fundamental variables. The formulas are extracted from the keplerian and classical gravitation laws, and  generally they contain a series of orbital parameters most of them achieved from experimental observations. So, it is not so easy to distinguish the exact role of each physical variable, even in the most rigorous procedures. In the present article, a series of calculations will be performed to express the orbital speed in terms of fundamental and physically observable variables. Principally, the results of a previous report will be used \cite{nota1}. In that note, the intensity field evolution was predicted at a given point moving at relative constant speed respect to an electromagnetic source. Taking into account the source emission mechanism, fundamentally a decreasing potential with the distance growth, and pointing out the geometrical parameters of the physical environment, the dynamics of the system was analytically described. As shown, the results  agree with the solutions obtained by the classical physical treatises,  for instance the retarded potential evaluation, despite the greater simplicity. Synthetically, in the cited note, the space-time scalar parameter which determines the potential energy,  the quantity $R$ (there, $dR$ called) does not correspond  simply to the real distance between the source and the observation point, but it is the sum of the real distance plus the distance between the same objects evaluated at the instant when the intensity field was null. In the present note, taking into account the analogy of  the electromagnetic and gravitational field, the same criterion will be employed. In summary, in the next sections, the basic statements of the previous procedure will be highlighted, then the analytical formula will be used to calculate the orbital speed, adopting as orbital parameters the potential energy of the system and the initial momentum of the moving object. The specific case of a constant energy potential will reveal a natural and strong connection with the elliptical trajectory.  In light of these statements, the trajectories and the orbital velocity of the solar system planets will be examined. 
 \section{Reference system and and fundamental parameters}
 \label{rif}
\begin{figure}
\includegraphics[width=7cm, angle=-90]{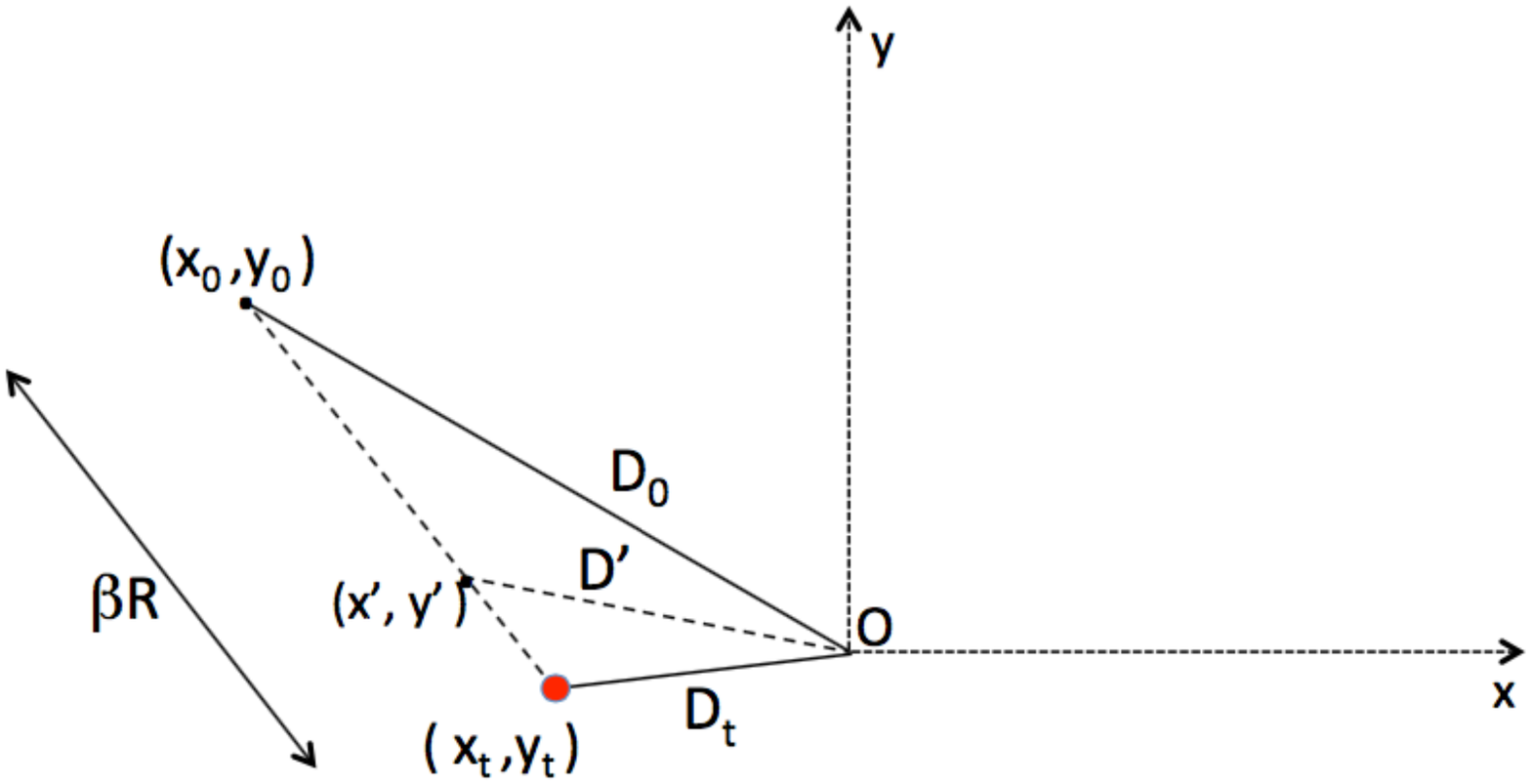}
\caption{
Space coordinates and other geometrical parameters relative a body (red bullet) bumping into a gravitational field at time $t$.
\label{space1} }
\end{figure}

Let's consider a body with mass $m$ at the position $(x_t,y_t)$ at the instant $t$, in a spatial reference system with origin into $ O$, as shown in fig\ref{space1}. {(Note: from now on, only a two-dimensional space will be considered)}. Initially, in absence of field, it moves at velocity $v_0$. Presuming the point $O$ as the center of the mass generating the gravitational field, the aim is to characterize the system at instant $t$, from the physical point of view.
As previously demonstrated \cite{nota1}, at that space-time coordinates, the intensity of a field propagating at light-speed $c$,  depends on $1/R^2$, where $R$  is the real path covered during the time interval $(t-t_0)$ \cite{nota1}, being $t_0$ such that 
\be
R=c(t-t_0)
\label{r0}
\ee
and
\be
\beta R=\sqrt{(x_{t}-x_{0})^2+(y_{t}-y_{0})^2}
\label{r_general}
\ee
where $\beta\equiv v_0/c$.
\begin{figure}
\includegraphics[width=7cm, angle=-90]{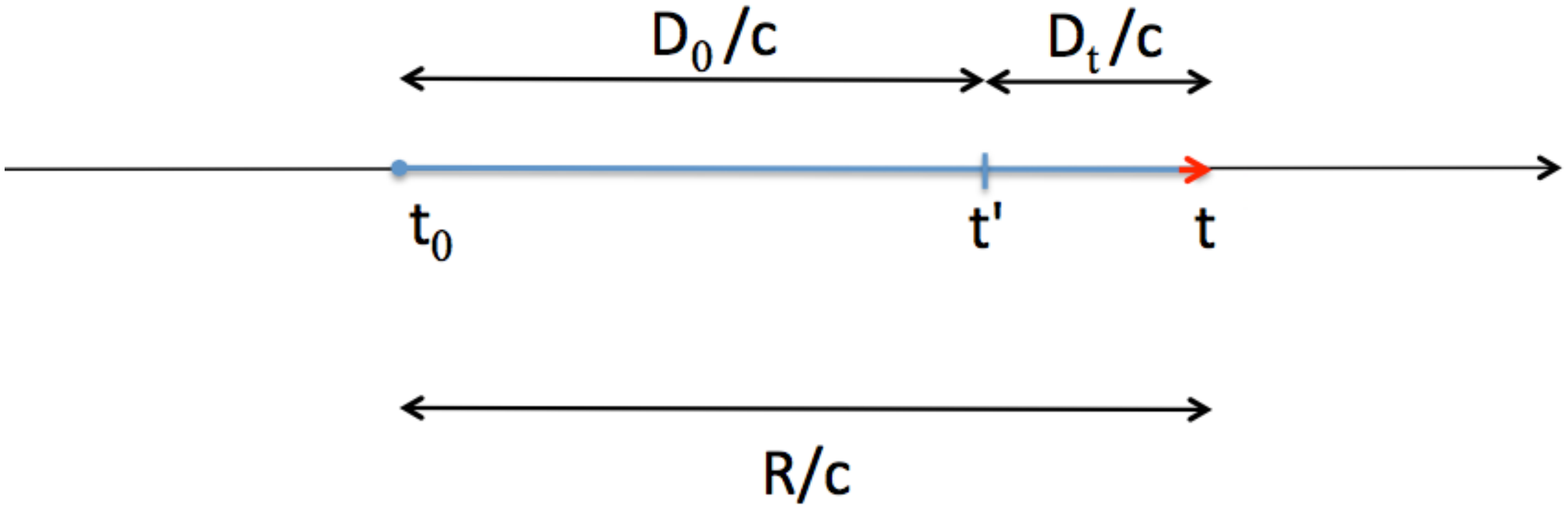}
\caption{The figure shows the time variables of the eqs.\ref{r0} and \ref{def_t}, and the time arrow. That one is pointed towards the right side with the red tip corresponding to the instant $t$.  
\label{time} }
\end{figure}
Considering the time parameter as an independent and linear variable, $t$ and $t_0$  must belong to the same time reference line (fig.\ref{time}). It is fundamental and useful to define an intermediate instant $t'$, on the same time line, in such a way
\be
t=t'+\frac{\sqrt{x_{t}^2+y_{t}^2}}{c}~~~~~~~{\text{and}}~~~~~~~~t_0=t'-\frac{\sqrt{x_{0}^2+y_{0}^2}}{c}
\label{def_t}
\ee
\begin{figure}
\includegraphics[width=7cm, angle=-90]{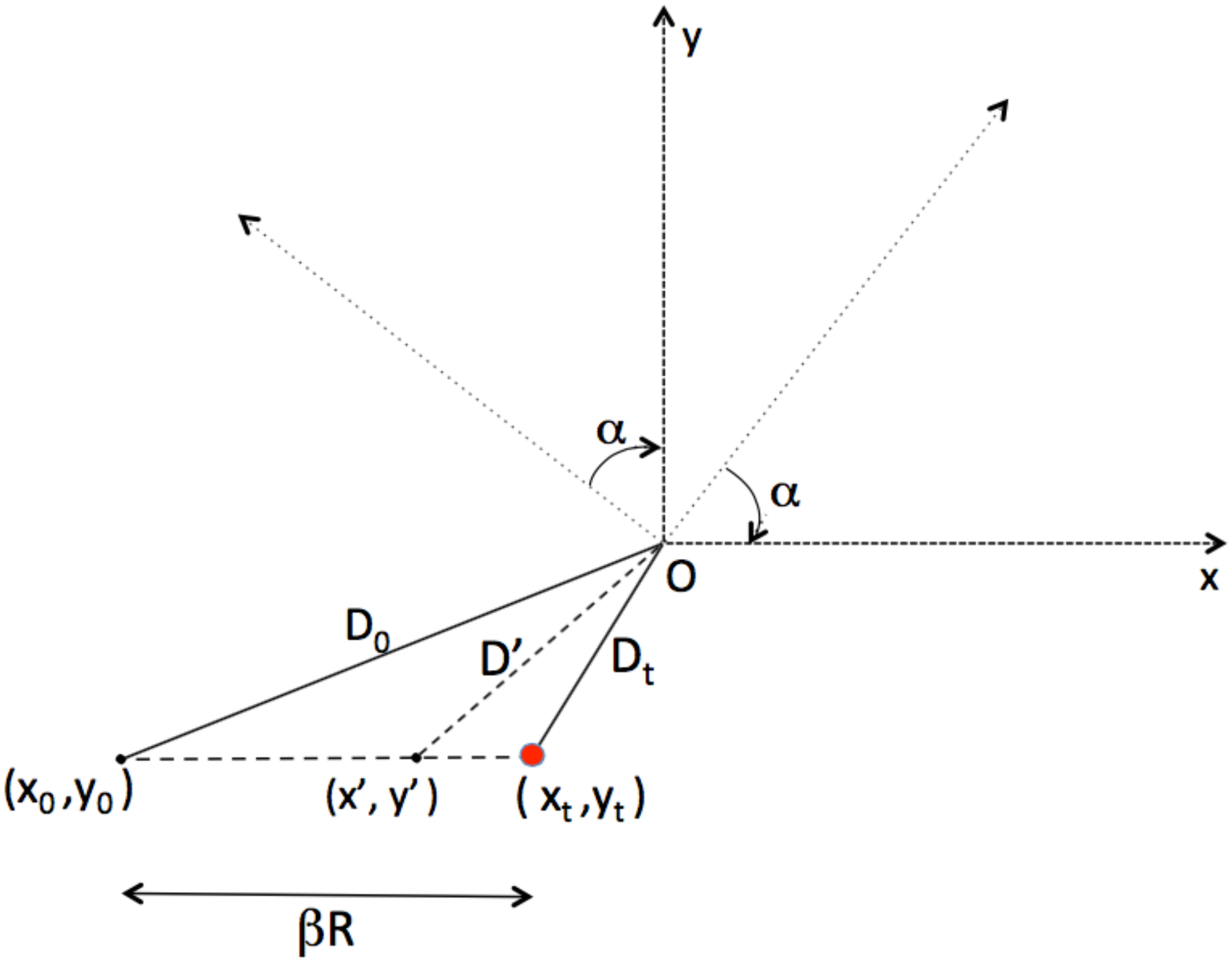}
\caption{
Rotating the reference system of fig\ref{space1} of an angle $\alpha$. Thus, the ordinate value at instant $t$ assumes the same value at time $t_0$, but the metric of the system is unchanged. The vector $\bf v_0$ becomes $[v_0,0]$. 
\label{space2} }
\end{figure}
At same time, in terms of space coordinates, we find $x'$, $y'$ and the relative length $D'\equiv\sqrt{x'^2+y'^2}$. 

Being $D_t\equiv\sqrt{x_t^2+y_t^2}$ and $D_0\equiv\sqrt{x_0^2+y_0^2}$ the distances of the object from the center point $O$ respectively at time $t$ and $t_0$, it results
\be
R=D_t+D_0.
\label{r_sum}
\ee

Rotating the reference system of an angle $\alpha$ so that $y_t=y_0$ (fig.\ref{space2}), the relation \ref{r_general} becomes 
\be
\beta R=|x_{t}-x_{0}|
\label{r_beta}
\ee
and the same length $R$ can be expressed also in the following ways
\be
 R=2\frac{(D_t-\beta x_t)}{1-\beta^2},
\label{r1}
\ee

\be
R=2\frac{(D_0+\beta x_0)}{1-\beta^2}.
\label{r2}
\ee

 \section{trajectory with  constant potential energy}
 \label{traj}
Meaning the trajectory as just the set of the covered spatial positions, the four eqs.\ref{r_sum}-\ref{r2} are enough for defining it. Also a reference system where $y_t=y'=y_0$ is necessary, and it is essential to underline that it is always possible to perform such a choice, without invalidating the metric properties of the system. That is easily understandable looking at the eqs.\ref{r_sum}-\ref{r2}, and considering the spatial transformation invariance of the quantities $R$, $D_t$ and $D_0$. So, regarding that equations, they describe the physical and spatial properties of the body of mass $m$, that runs into a gravitational field, just at instant $t$.  The relative potential energy is $E\sim 1/R$, and the momentum is $q=\beta mc$, so considering invariant that two quantities, means keeping constant $R$ and $\beta$. \\
In that conditions, for the body $m$, the locus $(x_t,y_t)$ of the allowed positions is an ellipse\\
- with one focus centered into $ O$\\
- with $~2a=R$, and\\
- with eccentricity $\epsilon=\beta$, being $a$ the major semi axis.\\
In a natural way, these statements reveal the geometrical properties of such a physical system. On the other hand, corresponding with the first keplerian law, they disclose the intrinsic physical meaning of an orbital path.
In that sense, trying to better understanding its physical properties, we can say that the eccentricity  indicates the specific momentum of the orbiting body, coinciding to $\beta=v_0/c$, where $v_0\equiv q/m$ can be interpreted as the velocity of the body respect to the gravity center, before it is trapped into the gravitational field. In fact, it is natural to perceive an initial high speed object generating a strongly elliptical track, and obviously vice-versa. Then, being  $R$ expression  of the potential energy, it can be invariant although it is the sum of the two variables (eq.\ref{r_sum}), just like the ellipse outline is the  locus which the sum of the distances to the two focal points is constant.

\begin{figure}
\includegraphics[width=1\linewidth]{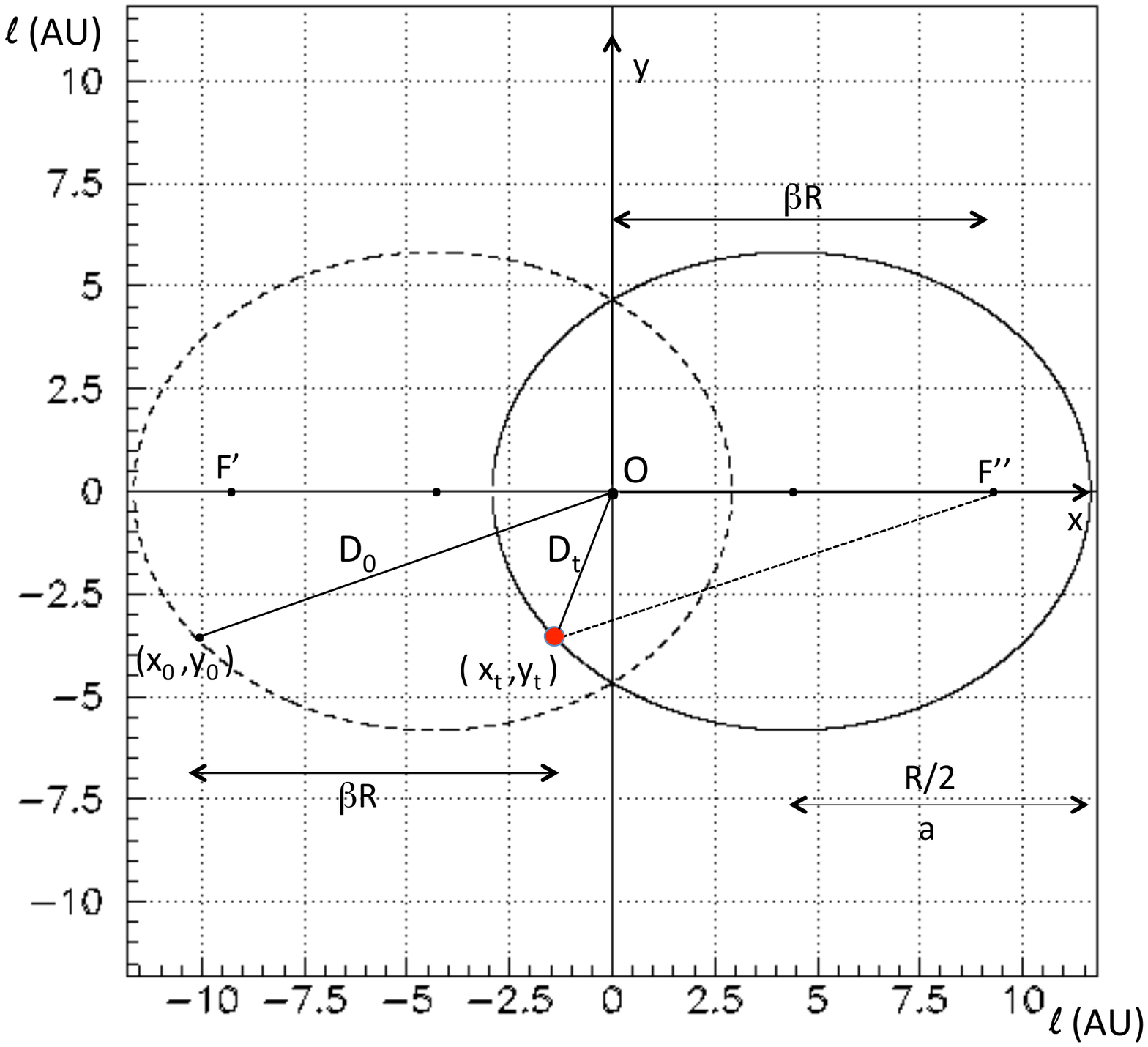}
\caption{
In the figure, the parameters solving the system equations \ref{r_sum}-\ref{r2} are indicated for a body with momentum $q=0.6~mc$, that is $\beta=0.6$, and $R=14.57$. The real orbit is described by the continuos line forming an ellipse mainly developing on the right side. It represents the only space locus $(x_t,y_t)$, for the defined reference system. On the left, the set of the points $(x_{0}, y_{0})$ is crosshatched. Both lines define two identical ellipses surrounding the focal points $F'$, $F''$, and $O$ in common, corresponding that to the origin of reference system, as well as to the center of the gravitational system. The semi axis length is $a=R/2$, and the eccentricity is $\epsilon=\beta=0.6$. A generic point on the trajectory is drawn with real distance $D_t$, and the relative length $D_0$.  
\label{tra1} }
\end{figure}

\begin{figure}
\includegraphics[width=1.\linewidth]{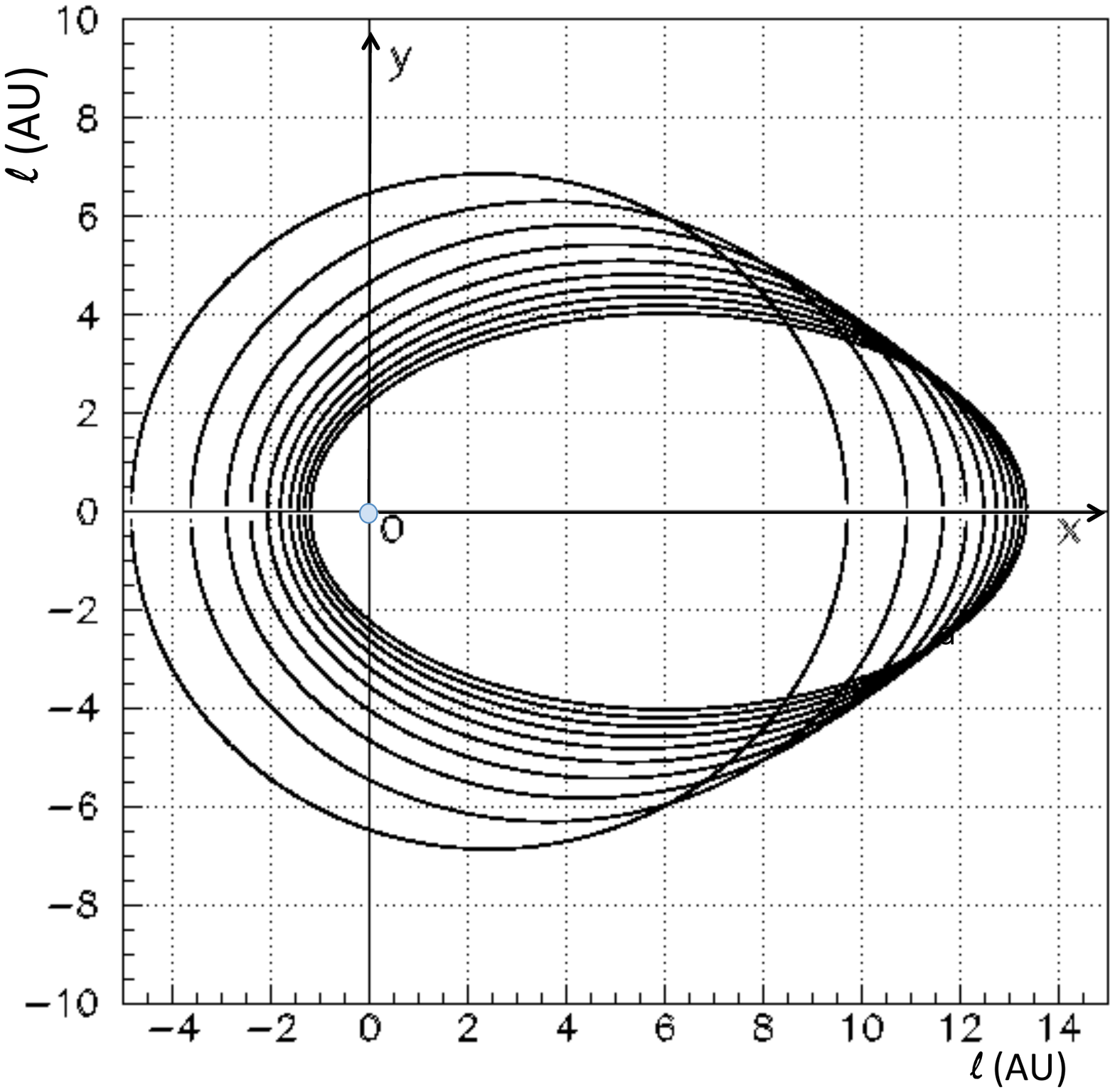}
\caption{
In a reference system in arbitrary unit (AU), some trajectories are shown, that are the couple sets $(x_t,y_t)$ solving the eqs.\ref{r_sum}-\ref{r2}, each one with the same length $R=14.57$, and with different $\beta$ values included into the interval $[0.33,0.83]$. The reference system is chosen in order to $v_0$ gets null contribute on the $y$ axis, i.e. $\bf v_0$$[v_x,0,]$. The elliptical figures are easily recognizable, all of that with $2a=R$, but different eccentricities $\epsilon$, each one corresponding to the fixed $\beta$ value. The curves flatten themselves while the $\beta$ value grows.
\label{tra10} }
\end{figure}

 Let's consider a practical example relative to a body having initial momentum $q=0.6mc$ that is $\beta=0.6$, and  $R=14.57$ in arbitrary units (AU). The relative elliptical trace is outlined in fig.\ref{tra1}. It represents the set of possible solutions of the equation system \ref{r_sum}-\ref{r2}. In the same figure, all the variables satisfying that equations are shown. In particular, the set of point $(x_0,y_0)$ corresponds to the crosshatched ellipse whose dimensions are exactly equal to the $(x_t,y_t)$ ellipse, with one focus anyway into $\bf O$. It is important noticing that also $(-x_0, \pm y_t)$ satisfies the eq.\ref{r1}, as though $(-x_t,\pm y_0)$ satisfies \ref{r2}, making so evident the complete overlapping of the two curves for less than a space-time phase quantity, and reaffirming so the stationary configuration of the system. 
In the fig.\ref{tra10}, there are ten closed orbits, each one with the same $R$ but different value for $\beta$. The curves flatten oneself  while the $\beta$ value grows. One of the two focal points is anyway steady in the origin of reference system, for each orbit.

\section{orbital velocity}
The orbital speed  expresses the dynamics of a body passing through a gravitational field, so it is characterized by the physical and spatial parameters, each time. 
 Trying to define $v_{orb}$, the following hypotheses are considered: the gravity is the only force, the total mass $M$ of the system is concentrated into $O$, so if $M_O$ is the mass at center, $M=m+M_O\approx M_O$. Then, taking into account the fundamental properties of the system, consecutive reasoning leads to following form\\
\be
v_{orb}(t)=\frac{D'}{D_t}\sqrt{\frac{2GM_O}{R(1-\beta^2)}}=\frac{D'}{D_t}\sqrt{\frac{2\gamma^2GM_O}{R}}
\label{v_orbi1}
\ee
where $G$ is the gravitational constant, and $\gamma\equiv 1/\sqrt{1-\beta^2}$.

\begin{figure}
\includegraphics[width=0.8\linewidth]{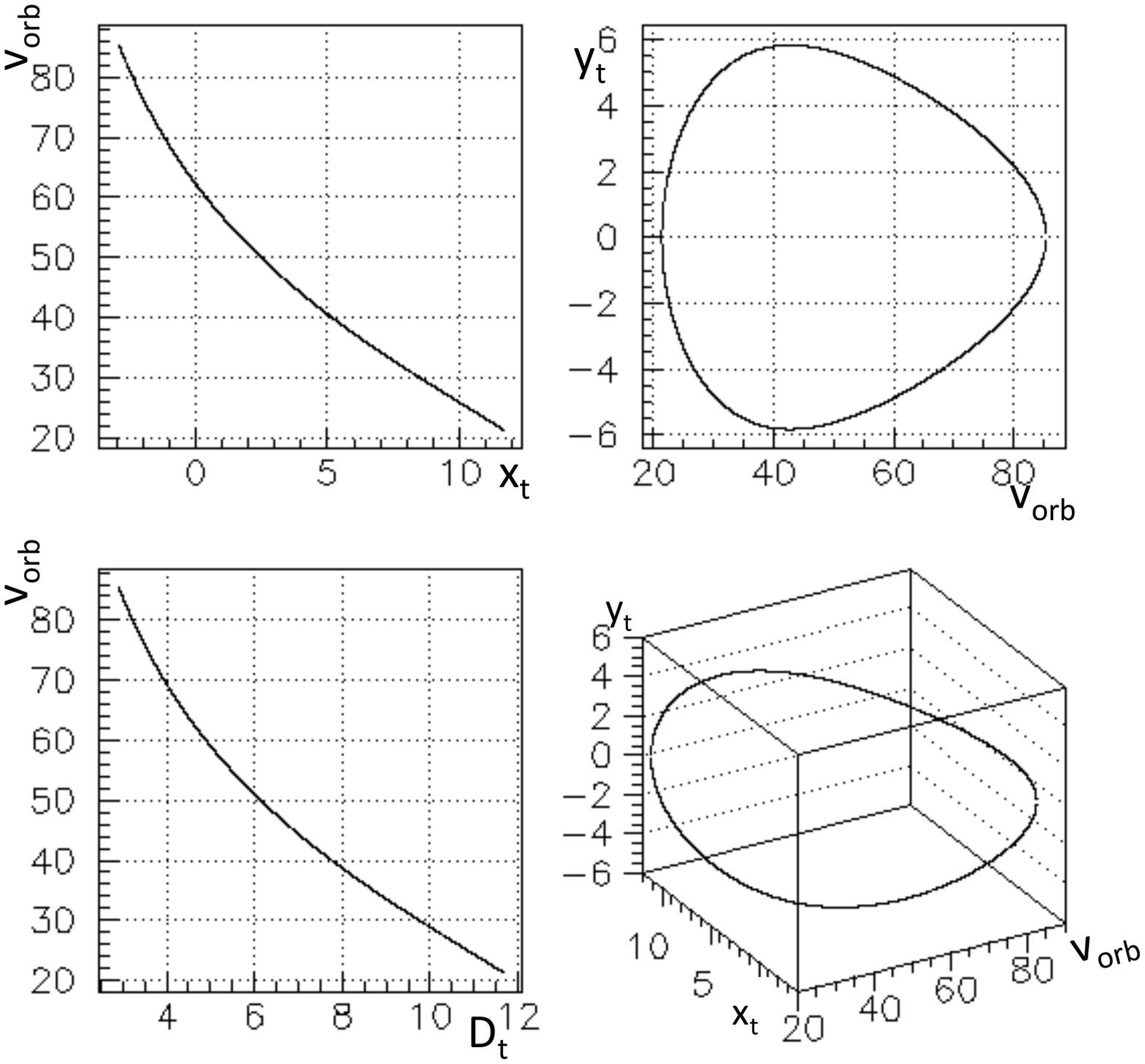}
\caption{The orbital velocity is reported as a function of different parameters, all of them in terms of AU. The calculation is related to the same example of fig.\ref{tra1}.
\label{speed1} }
\end{figure}

 The formula \ref{v_orbi1} can be rewritten for the specific case of a stationary orbit, that is $R=const=2a$, and choosing a special reference system like in fig.\ref{space2} where $x'= x_t-\beta D_t$. So, taking into the account the relation $D'= \sqrt{(x'^2+y_t^2)}$, the eq.\ref{v_orbi1} becomes
 \be
 v_{orb}(t)=\sqrt{\frac{\gamma^2 GM_O}{a}\left[1+\beta^2-\frac{2\beta x_t}{D_t}\right]}.
 \label{v_orbi2}
 \ee
 \\
 In fig.\ref{speed1}, the orbital velocity is reported in terms of the geometrical parameters. It has been calculated by eq.\ref{v_orbi2}, using the the same numerical values of fig.\ref{tra1}.\\
 
 Although  that quantity is obviously depending on time, the last two expressions do not contain an explicit temporal term. 
Effectively, tracing the trajectory, until now, the relative points have been considered as positions temporally independent each from the others, connecting each point to the real time $t$. In spite of the presence of several time quantities, only $t$ defines a real position of the object, while the values $t'$ and $t_0$ contribute to describe the physical state anyway concerning $t$. Thus, let's divide a body trajectory in $n$ points $(x_n,y_n)$. Each of them corresponds to the time $t=t_1,t_2,...,t_n$, entailing so a different system configuration with own $t'_1,t'_2,...,t'_n$, and $t_{01},t_{02},...,t_{0n}$ and the set related variables that satisfies all the previous equations.  Such a way, the $v_{orb}(t_n)$ value is univocally assigned to each point, by eq.\ref{v_orbi1}. 

With the aim to calculate the orbital time period, let's consider an elliptical trajectory. Coming after some logical steps, the following relation can be adopted 
\be
T_{orb}=\frac{\Pi_{el}}{\sqrt{8GM_OR}}
 \left( \gamma^{1/2}\sum_{i=2}^{n} |x_i-x_{i-1}|-\gamma^{2/3} P_{el} \right) 
\label{perio}
\ee
where $\Pi_{el}$ is the length of the orbit, $x_{i-1}$ and $x_i$ are consecutive abscissa coordinates related to the instants $t_{i-1}$ and $t_i$, and $P_{el}$ is the perihelion length. The relation is probably to improve with further study, since in the present report the approximation level is not well defined.
\section{ The solar system}
\label{sun}
For performing the experimental test on the previous formulas, the solar system is considered, since it is well conformed to the physical cases debated in the previous sections. Summarizing, that cases expect the trajectory that lies in a plane, a fixed center of mass $O$, and $M_0\gg m$, with $M_0$ equal to the solar mass, and $m$ corresponding to the single planet mass.
\subsection{Planet orbits}
\begin{figure}
\includegraphics[width=0.6\linewidth]{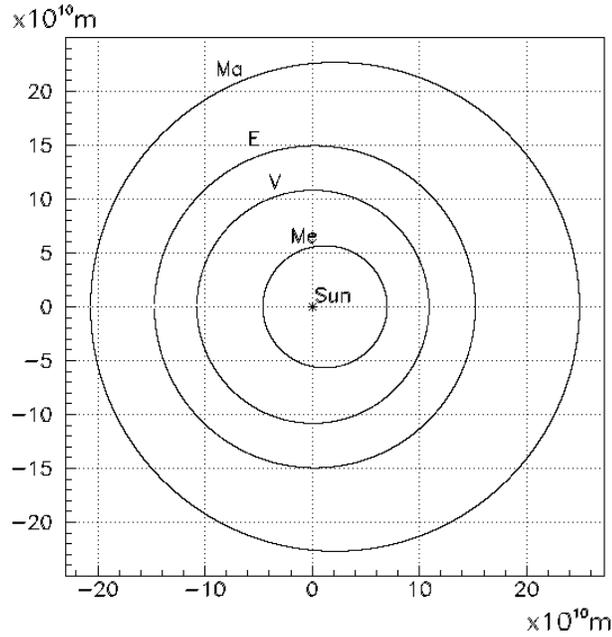}
\caption{The trajectories of the four planets nearest the Sun are drawn. They are been calculated taking into account the eqs.\ref{r_sum}-\ref{r2} and consequently the statements contained in the section \ref{rif} and \ref{traj}. The experimental  parameters $2a$ and $\epsilon$ are been extracted from the data tables in \cite{nasa},\cite{wiki}.
\label{solar1} }
\end{figure}

\begin{figure}

\includegraphics[width=0.6\linewidth]{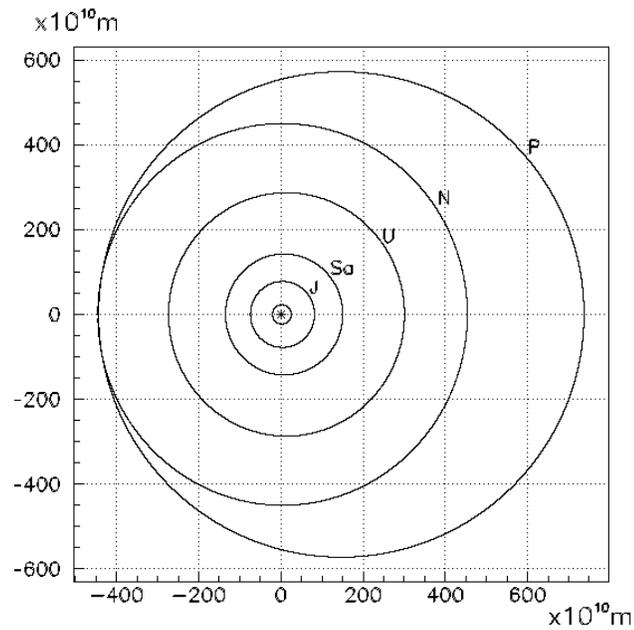}
\caption{The trajectories of the four planets more distant from Sun are shown, plus that of Pluto, the largest one. As in the previous figure, they derive from the statements contained in the sections \ref{rif} and \ref{traj}, and from data tables in \cite{nasa},\cite{wiki} regarding the ellipse dimensions.
\label{solar2} }
\end{figure}

Referring to the planets (including Pluto) of the solar system, let's trace  the nine relative orbits trying to find as much as couple possible set $(x,y)$ of the object positions. Essentially, two parameters are essential for each orbit, the major axis length $2a$ and the eccentricity $\epsilon$, two geometrical quantities to reveal from the data tables \cite{{nasa},{wiki}}. Assigning the physical  meaning as seen, that is $2a=R$ and $\epsilon=\beta$, these variables can be introduced into the eqs.\ref{r_sum}-\ref{r2}. The result is shown in fig.\ref{solar1} and \ref{solar2}. As predicted in the previous section, the resulting trajectories are elliptical and coinciding with the true planet orbits, with the Sun naturally centered in one focus, according to the keplerian law.
\begin{figure}
\includegraphics[width=1.\linewidth]{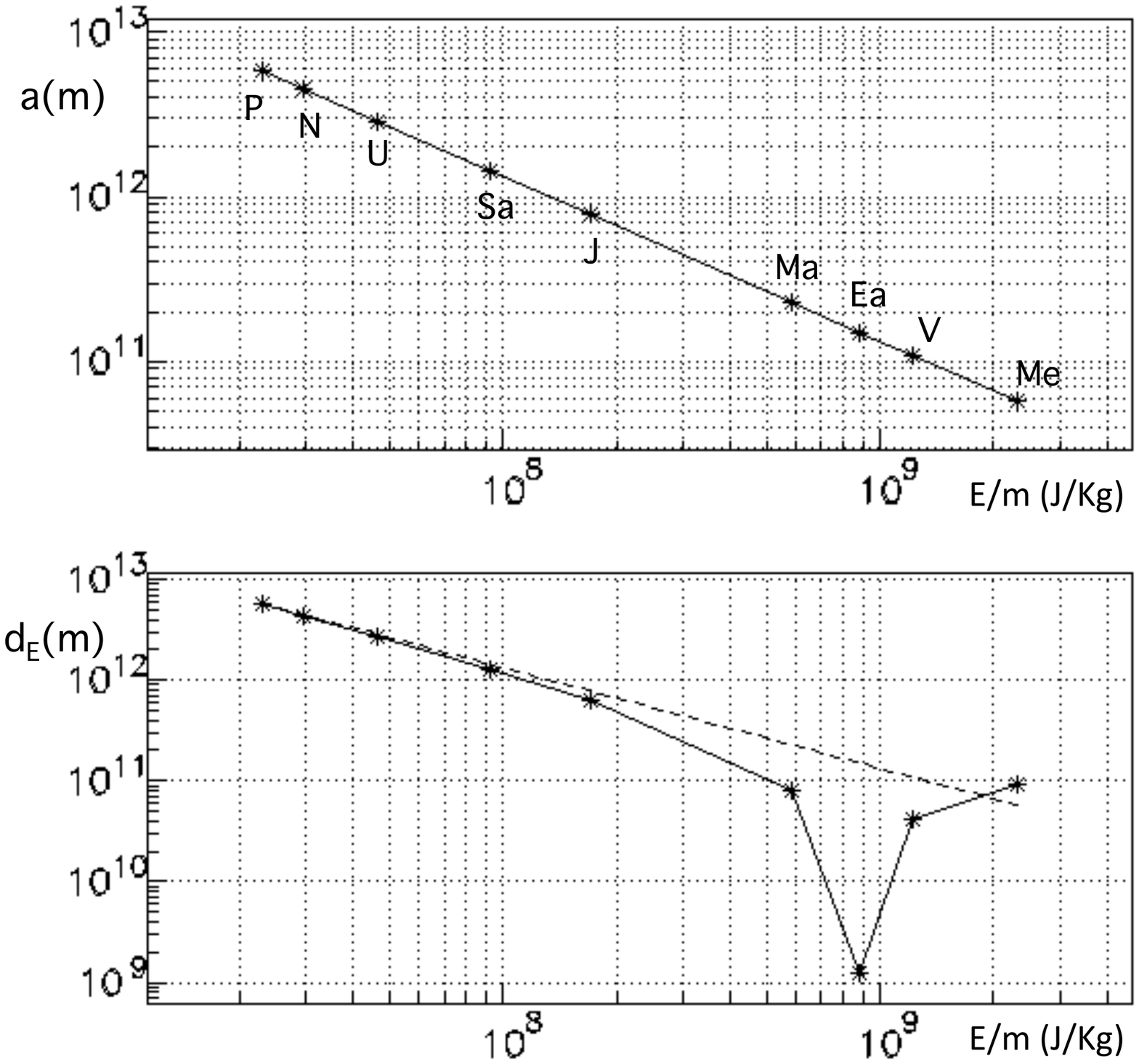}
\caption{ Spatial distribution for the orbiting objects in the solar system. In the upper graph, where the orbital semi axis length $a$ is reported, the observational point is the Sun. In the lower one, the average distances $d_E$ are evaluated from Earth. The crosshatched line is just the line of the upper graph, also there reported to improve the comparison. Remembering that $a$ is also a physical variable being $a=R/2$, while $d_E$ represents just a spatial distance, the discrepancy  between the two curves is better interpreted.
\label{terra} }
\end{figure}
\subsection{Orbital velocity and period}
The orbital speed is evaluated for each of the nine body belonging to the solar system. The eq.\ref{v_orbi2} is applied along the whole trajectories, tabulating the average with the minimum and maximum values in tab.\ref{tab1}. That two values correspond to the object positions relatively at aphelion and perihelion. The agreement with the measurements \cite{{nasa},{wiki}} is good especially for the minimum and maximum values that regard punctual body positions. The major discrepancy about the average value can be due to the different algorithms used for the average elaboration. In fig.\ref{terra}, regarding the same system, the specific energy spectrum is reported. The upper graph describes the distribution of the objects in terms of major semi axis $'a'$ relative to each own orbit, meaning each average distance from Sun.  The lower one 
concerns the same data but reporting the average distance $d_E$ from Earth, that is watching the same mass distribution but from another point of view. The comparison between the two curves (the lower graph shows both) is useful also from another perspective, considering that the variable $a$ is not simply a geometrical quantity but it contains an important physical significance being $a=R/2$, while $d_E$ is just a spatial distance. So, not giving sufficient consideration to that difference, the interpretation of results might be ambiguous or even erroneous.
The time period is calculated for each planet of the solar system and for Pluto, using the eq.\ref{perio}. Also here, there are some discrepancies with respect experimental values \cite{{nasa},{wiki}}, as reported in the last column of the tab.\ref{tab1}. Anyway, the real errors can be hardly calculated since they can due to different kind of normalization with respect the experimental data. Further, as said before, the eq.\ref{perio} probably must be better adapted, then the formula have been considered for a simplified system with just two bodies with $M_0\gg m$, without considering other perturbations. 
\begin{table}[]
\centering
\caption{Velocity values and orbital period for the planets of the solar system, including Pluto. The formula \ref{v_orbi2} is applied. The orbital period $T_{calc}$ is calculated by eq.\ref{perio}, while $T_{mea}$ is extracted principally from the NASA tables \cite{{nasa},{wiki}}.}
\label{tab1}
\begin{tabular}{c|clclclc|}
& & & &\\
 &$  v_{min}$   & $~~~~~\bar{v}$& $v_{max}$ &$~~~~$Orbital period  \\
 &$  (Km/s)$   & $~(Km/s)$& $~(Km/s)$ &$~~~~~~~ T_{calc}(days)$&$~~~T_{calc}/T_{mea} $\\
 \hline
 Mercury ~~~& 38.858 &~~~48.215  &  ~~58.976& ~~~~~~~~87.918&0.9994 \\
  Venus~~~& 34.784 & ~~~35.020 &  ~~35.258& ~~~~~~~224.706 &1.00002 \\
 Earth~~~& 29.291 & ~~~29.786 &~~30.286  & ~~~~~~~365.257&1.000002 \\
 Mars~~~& 21.972 & ~~~24.164 & ~~26.497 &~~~~~~~686.914&0.99993 \\
 Jupiter~~~ & 12.440 & ~~~13.063 & ~~13.705 &~~~~~~~4334.60&1.0003 \\
 Saturn~~~&9.138  & ~~~9.649 & ~~10.179 & ~~~~~~~10756.6& 1.00003\\
 Uranus~~~& 6.485 &  ~~~6.802& ~~7.128 & ~~~~~~~30702.1&0.9998 \\
 Neptune~~~& 5.385 &~~~5.432  & ~~5.478 & ~~~~~~~60227.4&1.00007 \\
 Pluto~~~& 3.676 & ~~~4.790 & ~~6.112 & ~~~~~~~90548.3&1.0004 \\
 \hline
\end{tabular}
\end{table}


\section{conclusions}
The orbital parameters have been determined for an object traveling in a gravitational field. Two scalar quantities are necessary, $R$ and $q$. The first one, $R$ defines the gravitational potential, being that depending on the rate $1/R$. It doesn't represent simply the distance between the orbiting object and the gravity center, it is the sum of the same real distance plus the distance at the time when $R$ was null.  That distance is virtual, in that sense it locates a precise instant but not an unique spatial point. The second parameter, $q=\beta mc $, is the momentum value relative to the orbiting body. The applied formulae have been previously and more generally determined considering the intensity  variation of an electromagnetic field, depending on the dynamic distance between the source and the observation point. In any case they have been derived from geometrical analysis of the physical environment, considering the space euclidean and the time linear. In the present article, the simple case of a fixed center of mass has been treated. Testing the agreement with the experimented physical formulae, the orbits of the solar system have been analyzed. The agreements with the experimental data is very satisfactory and encouraging for extending the same criterion to other physical structures.\\

{\large\bf Acknowledgements }\\

It is a great pleasure to thank Giorgio Fornetti for his precious support and for the certainly interesting discussions.

 \end{document}